\documentclass[journal=jacsat,manuscript=article]{achemso}

\usepackage{chemformula} 
\usepackage[T1]{fontenc} 

\usepackage{amsfonts}
\usepackage{mathrsfs}
\usepackage{amsmath}
\usepackage{color}
\usepackage{graphicx}
\usepackage{bm}
\usepackage{amssymb}
\usepackage{xspace}
\usepackage{epstopdf}
\usepackage{dcolumn}
\usepackage{longtable}
\usepackage{multirow}
\usepackage{hyperref}

\author{Hao Chen}
\affiliation{Department of Physics, University of Science and Technology of China, Hefei 230026, China}
\alsoaffiliation{Phonon Engineering Research Center of Jiangsu Province, Center for Quantum Transport and Thermal Energy Science, Institute of Physics Frontiers and Interdisciplinary Sciences, School of Physics and Technology, Nanjing Normal University, Nanjing 210023, China}

\author{Qianqian Wang}
\affiliation{Research Laboratory for Quantum Materials, Singapore University of Technology and Design, Singapore 487372, Singapore}

\author{Xukun Feng}
\affiliation{Research Laboratory for Quantum Materials, Singapore University of Technology and Design, Singapore 487372, Singapore}

\author{Weikang Wu}
\email{weikang_wu@sdu.edu.cn}
\affiliation{Key Laboratory for Liquid-Solid Structural Evolution and Processing of Materials, Ministry of Education, Shandong University, Jinan 250061, China}

\author{Lifa Zhang}
\email{phyzlf@njnu.edu.cn}
\affiliation{Phonon Engineering Research Center of Jiangsu Province, Center for Quantum Transport and Thermal Energy Science, Institute of Physics Frontiers and Interdisciplinary Sciences, School of Physics and Technology, Nanjing Normal University, Nanjing 210023, China}

\title[An \textsf{achemso} demo]
{Phonon chirality manipulation mechanism in TMD interlayer-sliding ferroelectrics}

\begin{document}

\begin{abstract}
As an ideal platform, both the theoretical prediction and first experimental verification of chiral phonons are based on transition-metal dichalcogenide materials. The manipulation of phonon chirality in these materials will have a profound impact on the study of chiral phonons. In this work, we utilize the sliding ferroelectric mechanism to study the phonon chirality manipulation mechanism in transition-metal dichalcogenide materials. Based on first-principles calculations, we study the different effects of interlayer sliding on the phonon properties in bilayer and four-layer MoS$_2$ sliding ferroelectrics. We find that sliding can regulate phonon chirality and Berry curvature, which further affects the phonon angular momentum and magnetization under a temperature gradient and the phonon Hall effect under a magnetic field. Our work connects two emerging fields and opens up a new route to manipulate phonon chirality in transition-metal dichalcogenide materials through the sliding ferroelectric mechanism.

\end{abstract}


In 2015, the concept of chiral phonons was first proposed in the monolayer transition-metal dichalcogenide (TMD) MoS$_2$ material~\cite{zhang2015chiral}. Phonons at the corner of the Brillouin zone (BZ) have well-defined chirality, and threefold rotational symmetry further endows these chiral phonons with quantized phonon pseudoangular momentum (PAM). In 2018, researchers first demonstrated the phonon chirality in WSe$_2$ using infrared ultrafast optical pump technology~\cite{zhu2018observation}. In this process, chiral phonons carrying PAM participate in the intervalley transition of electrons. As an ideal platform for chiral phonon research, many novel effects and phenomena related to chiral phonons in two-dimensional (2D) TMD materials have been discovered, such as entanglement of single photons and chiral phonons~\cite{chen2019entanglement}, brightening of the momentum-dark intervalley excitons~\cite{li2019momentum,li2019emerging,liu2020multipath}, nonlinear valley phonon scattering~\cite{liu2021nonlinear}, vibrational dichroism of chiral valley phonons~\cite{pan2023vibrational}, etc. If the chirality of phonons can be manipulated in these TMD materials, it will have extraordinary significance and impact on the study of chiral phonons.

Ferroelectricity~\cite{xu2013ferroelectric,martin2016thin} offers a possibility for realizing this idea. Especially for displacement-type ferroelectric materials~\cite{ding2017prediction,cui2018intercorrelated}, the switching of electric polarization corresponds to the displacement of certain atoms, which may have an impact on phonon chirality. However, the ferroelectricity of monolayer TMD materials has not been discovered thus far. Surprisingly, a unique ferroelectric mechanism was proposed in 2017---sliding ferroelectricity~\cite{li2017binary}. The 2D interlayer-sliding ferroelectric material is composed of a monolayer material without ferroelectric polarization, and ferroelectric order switching is caused by sliding between the upper and lower layers~\cite{li2017binary,wu2021sliding,miao2022direct,meng2022sliding,yang2023across,ji2023general}. The concept has received intense attention from researchers and was recently demonstrated experimentally by Yasuda et al.~\cite{yasuda2021stacking} and Stern et al.~\cite{vizner2021interfacial}.

In this work, we combine the concepts of sliding ferroelectrics and chiral phonons to study the manipulation effect of phonon chirality in TMD materials. In the bilayer MoS$_2$ sliding ferroelectric material, we find that the sliding only flips the phonon chirality in the $x$ and $y$ directions but maintains the phonon chirality in the $z$ direction. In the four-layer MoS$_2$ sliding ferroelectric, the phonon chirality in all the $x$, $y$ and $z$ directions will flip with the interlayer-sliding. The modulation of chiral phonons by interlayer sliding also brings other interesting effects, such as the reversal of the total phonon angular momentum and magnetization under a temperature gradient and the reversal of the phonon PAM at high-symmetry points $K$ and $K'$. At the same time, in the four-layer MoS$_2$ sliding ferroelectric system, the switching of ferroelectric states changes the sign of the phonon Berry curvature, which also leads to a flipping of the phonon Hall conductivity.

\textit{\textbf{Phonon chirality and pseudoangular momentum}}. Phonon chirality is characterized by phonon polarization. A phonon polarization greater than zero indicates that the phonon is right-handed, while a phonon polarization less than zero indicates that the phonon is left-handed. For polar structures such as ferroelectric materials, phonon polarization appears not only in the $z$ direction but also in the $x$ and $y$ directions. Therefore, the expression of phonon polarization is:
\begin{eqnarray}\label{eq:polarization}
	s_{\hat{n}} = \hbar\psi^{\dag}\hat{S}_{\hat{n}}\psi
\end{eqnarray}
where $\psi$ is the phonon mode, $\hat{S}_{\hat{n}}$ is the phonon circular polarization operator, and $\hat{n}=x,y,z$.

The $z$-direction phonon polarization operator can be defined as~\cite{zhang2015chiral}
	\begin{equation}
		\hat{S}_{z}\equiv\sum_{\alpha=1}^n(|R_{z,\alpha}\rangle\langle R_{z,\alpha}|-|L_{z,\alpha}\rangle\langle L_{z,\alpha}|)=
		\left(
		\begin{array}{ccc}
			0 & -i & 0 \\
			i &  0 & 0 \\
			0 &  0 & 0
		\end{array}
		\right)
		\otimes I_{n\times n}.
	\end{equation}
Similarly, we can define phonon polarization operators in the $x$ and $y$ directions:
\begin{equation}
	\hat{S}_{y}=
	\left(
	\begin{array}{ccc}
		0 &  0 & i \\
		0 &  0 & 0 \\
		-i &  0 & 0
	\end{array}
	\right)
	\otimes I_{n\times n},\
	\hat{S}_{x}=
	\left(
	\begin{array}{ccc}
		0 &  0 & 0 \\
		0 &  0 & -i \\
		0 &  i & 0
	\end{array}
	\right)
	\otimes I_{n\times n}.
\end{equation}
The phonon eigenvector used in the calculation is obtained by VASP.

Additionally, after obtaining the phonon wave function, we can calculate the phonon PAM $\ell_{\text{ph}}$ at the high-symmetry points $K$ and $K'$. The phonon PAM corresponds to the phase difference after the threefold rotation operator is applied to the phonon mode. The formula is~\cite{zhang2015chiral}:
\begin{equation}\label{PAM}
	\mathcal{R}[(2\pi/3),z]\psi_{\bm q}= e^{-i(2\pi/3)\ell_{\text{ph}}^{\bm q}} \psi_{\bm q},
\end{equation}
where $\bm q=K,K'$ for the high-symmetry points here. The phase difference comes from the intercell part and intracell part. One can obtain orbital PAM $\ell_o$ for the intercell part and spin PAM $\ell_s$ for the intracell part.

\begin{figure}[htbp]
	\includegraphics[width=1\linewidth]{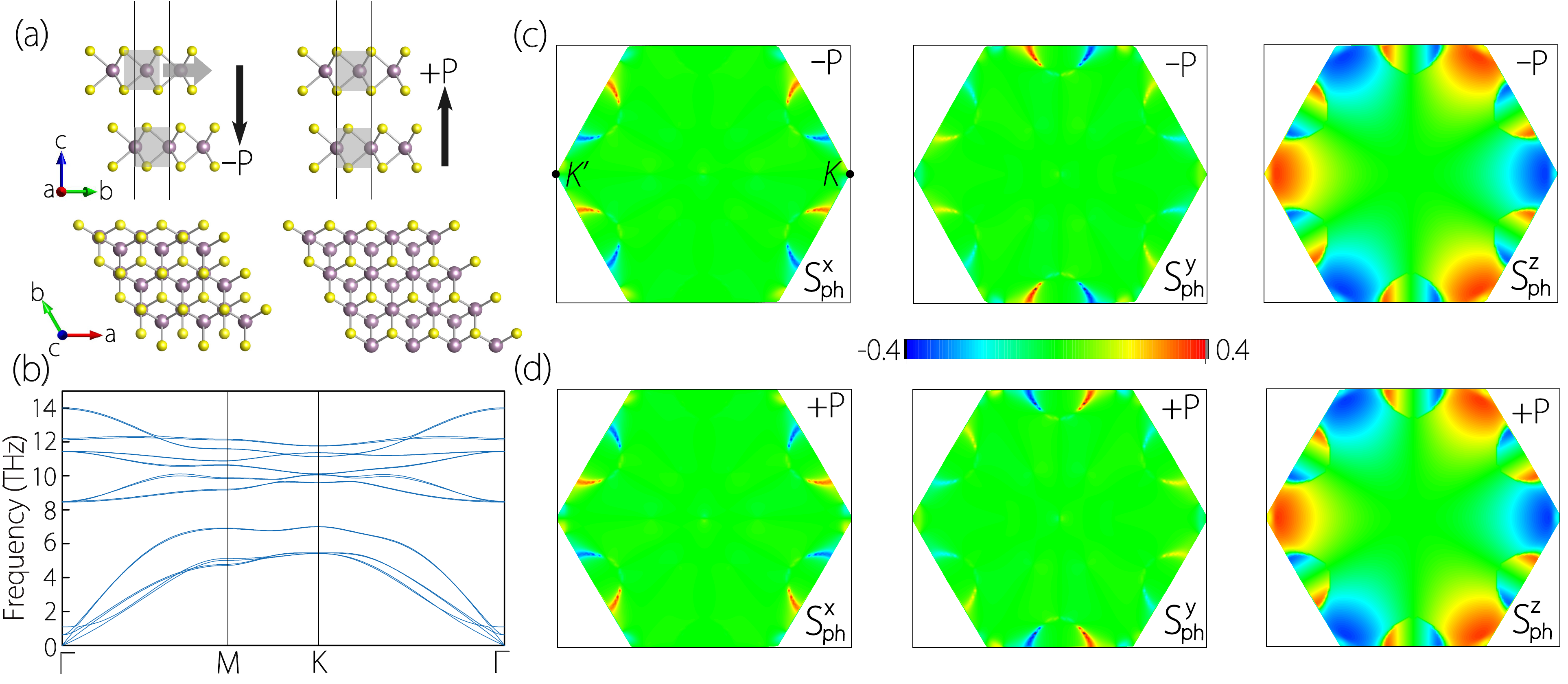}
	\caption{\label{fig1} Calculation results of the bilayer MoS$_{2}$ sliding ferroelectric. (a) The crystal structures of the $-P$ and $+P$ ferroelectric states ((upper panel: side view, lower panel: top view). The definitions of the $-P$ and $+P$ states are taken from the citation~\cite{wang2022interfacial,xiao2022non}. (b) The $-P$ and $+P$ states have the same phonon spectrum. (c) and (d) show the phonon polarization of the $-P$ and $+P$ states, respectively. Here, we take the first phonon branch as an example.}
\end{figure}

\begin{figure*}[htbp]
	\includegraphics[width=1\linewidth]{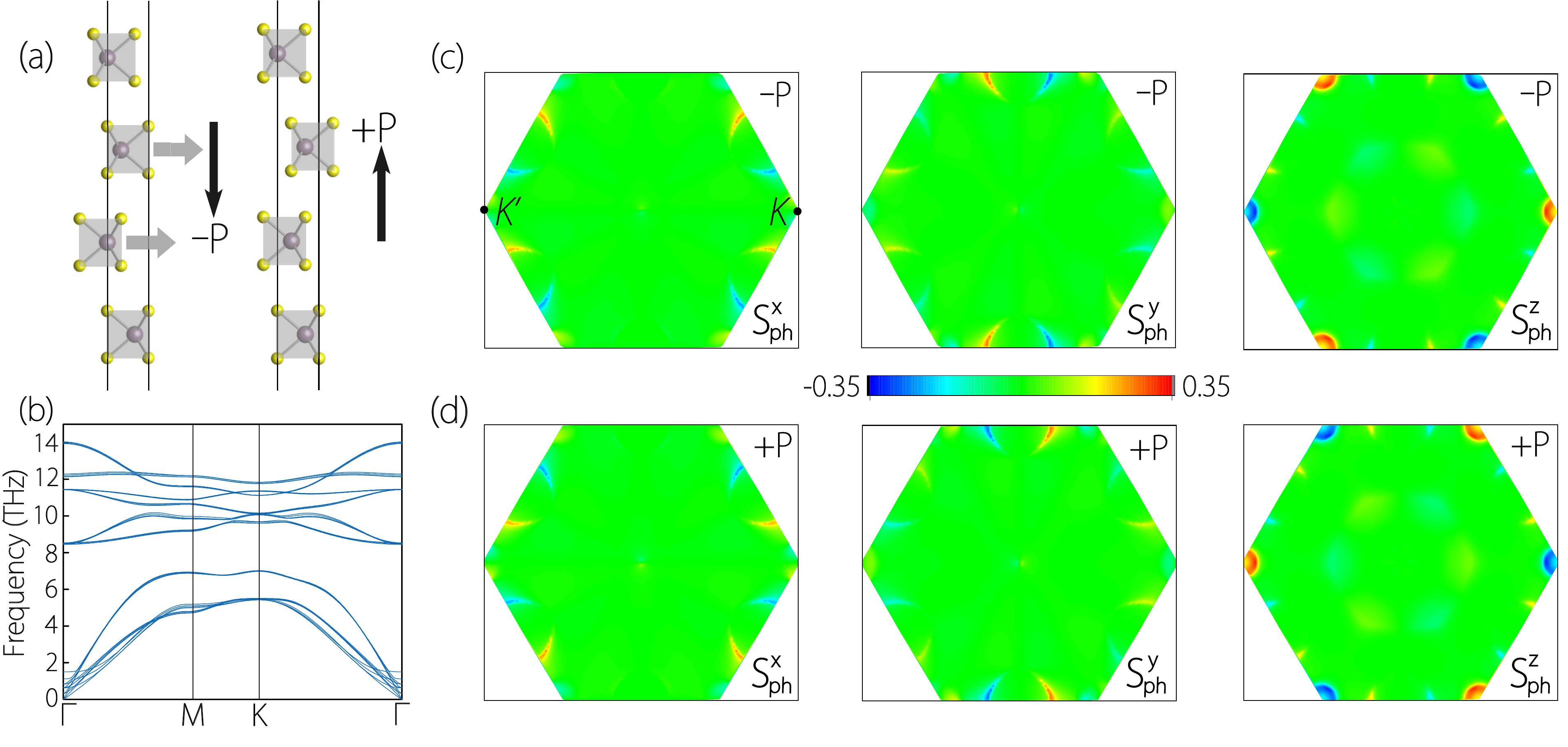}
	\caption{\label{fig2} Calculation results of the four-layer MoS$_{2}$ sliding ferroelectric. (a) Crystal structures of the opposite ferroelectric states; the definitions of the $-P$ and $+P$ states are taken from citation~\cite{liang2021intercorrelated}. (b) The $-P$ and $+P$ states have the same phonon spectrum. (c)-(d) Phonon polarization distributions in all three directions of the $-P$ state and $+P$ state, respectively. Here, we take the first phonon branch as an example.}
\end{figure*}

\textit{\textbf{Bilayer sliding ferroelectrics}}. Bilayer MoS$_{2}$ consisting of two identical nonferroelectric monolayers exhibits out-of-plane ferroelectricity and has been experimentally verified~\cite{wang2022interfacial}. Fig.~\ref{fig1} (a) shows the bilayer MoS$_{2}$ with the $-P$ and $+P$ ferroelectric states, and the switching of states can be realized by interlayer sliding. Their stacking ways correspond to AC and AB, respectively. For stacking details, please see the Supporting Information. The phonon spectra of the two states are the same, as shown in Fig.~\ref{fig1} (b). The unit cell has six atoms, so there are eighteen phonon branches, including three acoustic branches and fifteen optical branches. In addition, since each layer in the bilayer structure is the same MoS$_{2}$ monolayer, every two phonon branches in the spectrum have similar frequencies. Although two structures with opposite ferroelectric states $P$ have the same phonon spectrum, the corresponding phonon polarization has a different distribution. For bilayer MoS$_{2}$, only the phonon polarization in the $x$ and $y$ directions will be reversed after $P$ flipping, and the phonon polarization in the $z$ direction will not change, as shown in Fig.~\ref{fig2} (c) and (d). This is because the structures of the $-P$ and $+P$ states are connected by the $M_z$ symmetry, and the $M_{z}$ symmetry will only flip the wave vector $k_{z}$ (since it is a 2D material, there is no need to consider it) but will not change the wave vectors $k_{x}$ and $k_{y}$. Meanwhile, the $M_z$ symmetry will flip $s_\text{ph}^x$ and $s_\text{ph}^y$, but it has no effect on $s_\text{ph}^z$, which is perpendicular to the $z$ mirror plane. Therefore, for this 2D material, ferroelectric state switching connected by $M_z$ symmetry only flips the phonon polarization in the $x$ and $y$ directions.

Due to the $C_{3z}$ symmetry of bilayer MoS$_{2}$ (space group is $P3m1$), it can also define the phonon PAM at the high-symmetry $K/K'$ points. However, the PAM does not change with the flipping of the ferroelectric state $P$. Specifically, the invariance of phonon chirality in the $z$ direction corresponds to the invariance of spin PAM $\ell_s$. The two structures with opposite $P$ are connected by $M_z$ symmetry, which also means that from a top view, the Bloch phases corresponding to the sublattices are the same. They also have the same orbital PAM $\ell_o$. Therefore, the ferroelectric order switching in bilayer MoS$_2$ cannot flip the phonon PAM. For the calculated phonon PAM, please see the Supporting Information.

\textit{\textbf{Four-layer sliding ferroelectrics}}. Next, we turn our attention to the four-layer MoS$_2$ sliding ferroelectric, which can be seen as a double bilayer MoS$_2$. From fig~\ref{fig2} (a), we can see that due to the sliding of the middle two layers, the electric polarization of the material can be flipped from the $-P$ state to the $+P$ state. The stacking ways of these two structures are ABB$^{'}$A$^{'}$ and ACC$^{'}$A$^{'}$, respectively. From citation~\cite{liang2021intercorrelated}, we know that this stacking way has the lowest energy for a four-layer structure. They have the same phonon spectrum, as shown in fig~\ref{fig2} (b). Similar to bilayer MoS$_2$, four-layer MoS$_2$ has a similar frequency distribution of every four phonon branches.

To explore the manipulation effect of phonon chirality, we characterized the polarization distribution in the Brillouin zone. It can be clearly seen from fig~\ref{fig2} (c) and (d) that the phonon polarizations in all three directions are one-to-one correspondingly opposite. This is quite different from the results in bilayer MoS$_2$. At the same time, the phonon PAM at the high-symmetry points $K$ and $K'$ will also be flipped. The switching of phonon polarization in the $z$ direction directly corresponds to the reversal of spin PAM $\ell_s$. The reversal of orbital PAM $\ell_o$ is due to the spatial inversion distribution of the sublattice positions between two states. The calculated phonon PAM is shown in the Supporting Information.

We analyze the reason why ferroelectric state switching in four-layer MoS$_2$ can flip the phonon chirality in all three directions. Two opposite ferroelectric states can be connected by spatial inversion $\mathcal P$. Since each ferroelectric state respects time-reversal symmetry $\mathcal T$, it can also be said that two opposite states are related to each other by the combined $\mathcal {PT}$ operation. $\mathcal {PT}$ flips the phonon polarization $s_\text{ph}^x$, $s_\text{ph}^y$, $s_\text{ph}^z$ but preserves the wave vector $\bm k$. Therefore, for the same $\bm k$ of the two ferroelectric states, the phonon polarizations in all three directions are opposite.

\begin{figure*}[htbp]
	\includegraphics[width=4.8 in,angle=0]{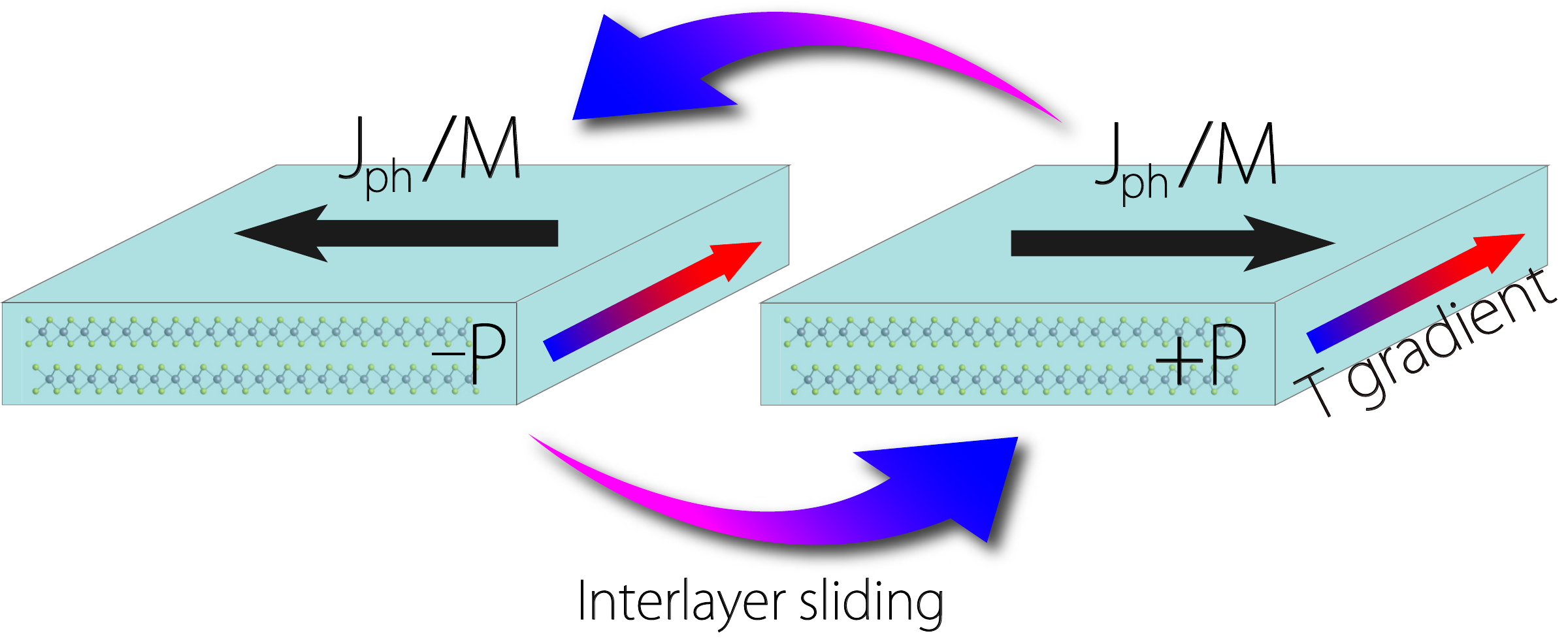}
	\caption{\label{fig3} Schematic diagram of phonon angular momentum and magnetization under a temperature gradient that can be controlled by interlayer sliding. $-P$ and $+P$ indicate the ferroelectric states. $J_{\text{ph}}$ represents the phonon angular momentum, and $M$ represents the phonon magnetization.}
\end{figure*}

\textit{\textbf{Phonon angular momentum and magnetization}}. For the nonmagnetic bilayer and four-layer MoS$_2$ structures, the $\mathcal P$ symmetry of the system is broken, while the $\mathcal T$ symmetry is maintained. Phonon polarization in different directions occurs in the Brillouin zone but is an odd function under $\mathcal T$ symmetry. Therefore, there is no total phonon angular momentum in the equilibrium state. However, when the temperature gradient is applied, the system will have total phonon angular momentum due to the unbalanced distribution of chiral phonons~\cite{hamada2018phonon},
\begin{equation}\label{J_temp}
	J_{\text{ph}}^{\hat{n}}=-\frac{\tau}{V}\sum\limits_{\bm{k},\sigma}s_{\sigma,\hat{n}}v_{\sigma,\hat{m}}\frac{\partial f_{0}(\omega_{\sigma}(\bm{k}))}{\partial T}\frac{\partial T}{\partial x_{\hat{m}}}=\alpha_{\hat{n}\hat{m}}\frac{\partial T}{\partial x_{\hat{m}}},
\end{equation}
where $\alpha_{\hat{n}\hat{m}}$ is the response tensor. For the bilayer and four-layer MoS$_2$ structures with the $C_{3v}$ point group, the response tensor matrix under the temperature gradient is:
$\alpha_{\hat{n}\hat{m}}=\left(
\begin{array}{ccc}
	0& \alpha_{xy} & 0 \\
	\alpha_{yx}& 0 & 0 \\
	0& 0 & 0
\end{array}
\right )$.
From our calculations, $\alpha_{xy}=-\alpha_{yx}\sim 10^{-7}\times[\tau/(1\text{s})]\,\text{Jsm}^{-2}\text{K}^{-1}$ at $T=300\, \text{K}$. From these matrix elements, we can see that it is only related to the phonon polarization in the $x$ and $y$ directions. As we mentioned above, for the bilayer and four-layer MoS$_2$ sliding ferroelectric, the phonon polarization in the $x$ and $y$ directions is completely reversed before and after the ferroelectric flip. Thus, the value of the matrix elements $\alpha_{xy}$ and $\alpha_{yx}$ will also be flipped. This means that interlayer sliding can regulate the direction of the total phonon angular momentum $J_{\text{ph}}$ under the temperature gradient.

Since the oscillating ions are charged, this induced phonon angular momentum is usually accompanied by a magnetic moment. Using the Born effective charge, we can estimate it. The relationship between the magnetic moment $\bm m$ and angular momentum $\bm j$ is $\bm{m}=\gamma\bm{j}$, where $\gamma$ is the gyromagnetic ratio. The gyromagnetic ratio tensors of the Mo and S atoms are given by $\gamma_{\alpha\beta}^{\text{Mo}}=geZ_{\alpha\beta}^{*}/2m_{\text{Mo}}$ and $\gamma_{\alpha\beta}^{\text{S}}=-geZ_{\alpha\beta}^{*}/2m_{\text{S}}$. The Born effective charge tensor $Z_{\alpha\beta}^{*}$ is calculated from VASP, where $m_{\text{Mo}}$ and $m_{\text{S}}$ are the masses of the Mo atom and S atom, respectively. Assuming that the g factor is on the scale of $1-10$, we roughly estimate that the magnitude of the magnetization is
\begin{equation}
	M\sim-\frac{\Delta T/(1\text{K})}{L/(1\text{m})}\times 10^{-11} \text{A/m}.
\end{equation}

Using the magneto-optical Kerr technique, $10^{-9}\mu_{B}/\text{nm}^{3} = 9.2732\times 10^{-6} \text{A/m}$ can be measured. To detect the phonon magnetization, a temperature gradient greater than $10^{5} \text{K/m}$ is needed. Under the current experimental conditions, a temperature gradient of $10^{6}-10^{7} \text{K/m}$ can be achieved~\cite{kim2023chiral}. Therefore, this phonon magnetization is expected to be detected experimentally. In these interlayer-sliding ferroelectric materials, the switching of the ferroelectric state $P$ is accompanied by the reversal of the phonon magnetization. The detection of this reversal could verify our idea.

\begin{figure}[htbp]
	\includegraphics[width=4.0 in,angle=0]{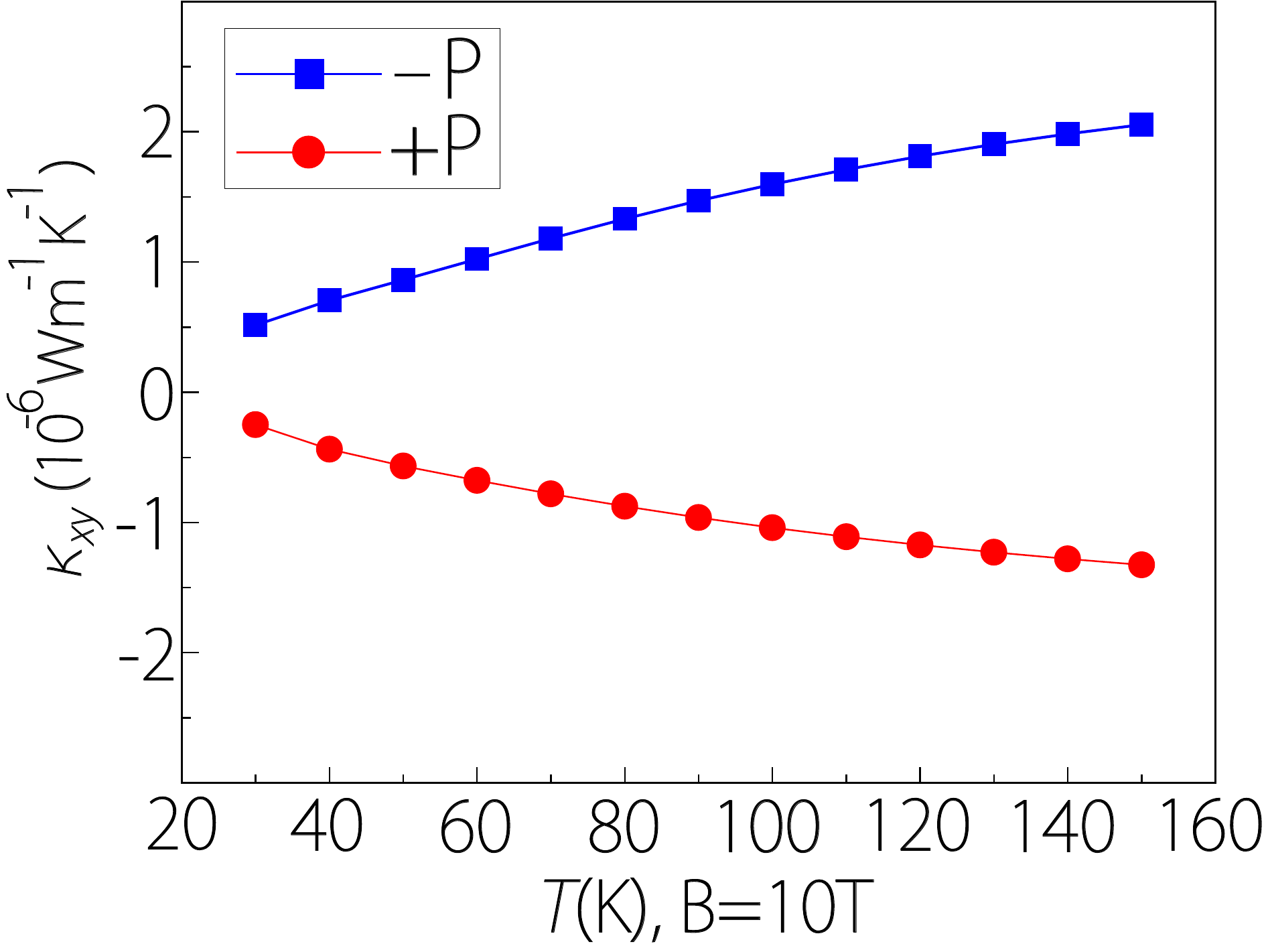}
	\caption{\label{fig4} Calculated phonon Hall conductivity $\kappa_{xy}$ versus temperature at B = 10 T. The blue squares represent the $-P$ state four-layer MoS$_2$, and the red balls represent the $+P$ state four-layer MoS$_2$.}
\end{figure}

\textit{\textbf{Phonon Berry curvature and Hall effect}}. In addition to phonon chirality, another physical quantity flipped during ferroelectric state switching is the phonon Berry curvature. For these 2D systems, the phonon Berry curvature here has only an out-of-plane component. Similar to the response to phonon polarization, only four-layer MoS$_2$ can switch the phonon Berry curvature in the $z$ direction. The calculated Berry curvature is shown in the Supporting Information. Because the phonon Hall effect is closely related to the Berry curvature, its switching may cause the flipping of the phonon Hall effect. The phonon Hall effect was first observed experimentally in 2005~\cite{strohm2005phenomenological}. After applying an external magnetic field, the nonzero phonon Berry curvature of the system will generate a transverse heat current under the longitudinal temperature gradient. Here, we demonstrate our idea through the calculation method of Sun et al.~\cite{PhysRevB.103.214301}. We focus on the self-consistent phonon (SCPH) approach~\cite{werthamer1970self,souvatzis2008entropy,errea2014anharmonic} and borrow the ALAMODE package~\cite{tadano2015self} developed by Tadano and Shinji.

The expression for the phonon Hall thermal conductivity is~\cite{qin2012berry,zhang2016berry,PhysRevB.103.214301}
\begin{equation}\label{kappa}
	\kappa_{xy}=\frac{1}{2\hbar T}\int^{\infty}_{-\infty} d\epsilon\ \epsilon^{2}\eta(\epsilon)\frac{\partial f_0(\epsilon)}{\partial\epsilon},
\end{equation}
where
\begin{equation}
	\eta(\epsilon)=\frac{1}{V}\sum_{n,\bm q}\Omega_{n\bm{q}}^{z}\Theta(\epsilon-\hbar\omega_{n\bm q}),
\end{equation}
and $\Theta$ is the step function, and $\Omega_{n\bm{q}}^{z}$ is the $z$ component of the Berry curvature.

The effect of the magnetic field on the system is introduced through the spin-phonon interaction. The form of it is
\begin{equation}\label{spi}
	H_I=\sum\limits_{i} \bm h_{\alpha}\cdot(\bm u_{\alpha}\times \bm p_{\alpha}).
\end{equation}
Assuming that the magnetic field is along the $z$ direction, the spin-phonon interaction can be written as
\begin{equation}\label{spi}
	H_I=u^{T}A \,p.
\end{equation}
where the $A$ matrix is~\cite{PhysRevB.103.214301}
\begin{equation}\label{spi}
	\bm A=\frac{e}{4\bm M_{\alpha}}(\bm Z_{\alpha}^{T}\times \bm B+\bm B\times \bm Z_{\alpha}).
\end{equation}
where $\bm Z_{\alpha}$ is the Born effective charge dyadic and $\bm B$ is the magnetic field.

Assuming a magnetic field of 10 T in the $+z$ direction is applied to the sample, the results show that the $-P$ and $+P$ states of the four-layer MoS$_2$ have opposite Hall conductivity values, as shown in fig~\ref{fig4}. This means that the interlayer sliding can flip the direction of the phonon Hall current. We notice that the absolute values of $\kappa_{xy}$ of the two states are not exactly equal, which is caused by the breaking of $\mathcal T$ symmetry. However, since the magnetic field has a weak influence on the phonon system, the difference is small. In the experiment, a phonon Hall conductivity on the order of $10^{-6}\text{Wm}^{-1}\text{K}^{-1}$ was measured~\cite{strohm2005phenomenological}. Therefore, the reversal of the phonon Hall effect is expected to be observable.

\textit{\textbf{Discussion and Conclusion}}. In this work, our research is based on the structures stacked from monolayer TMD MoS$_2$. It is worth emphasizing that these conclusions are not limited to MoS$_2$ but can be extended to all monolayer materials with $D_{3h}$ symmetry, such as TMDs, group \uppercase\expandafter{\romannumeral2} oxides, group \uppercase\expandafter{\romannumeral3}-\uppercase\expandafter{\romannumeral5}/\uppercase\expandafter{\romannumeral6} monolayers, etc. In these noncentrosymmetric monolayers, the $\mathcal P$ symmetry is broken. In addition, this mechanism is also applicable to monolayer materials with $\mathcal P$ symmetry. We discuss the trilayer PtSe$_2$ interlayer-sliding ferroelectric in the Supporting Information. The single-layer PtSe$_2$ has $D_{3d}$ symmetry, which maintains $\mathcal P$. Since two opposite ferroelectric states of trilayer PtSe$_2$ are connected by $\mathcal P$ symmetry, the ferroelectric state switching caused by sliding can flip the phonon polarization in all three directions.

In conclusion, we investigate the phonon chirality manipulation mechanism in TMD interlayer-sliding ferroelectric materials. For bilayer and four-layer MoS$_2$, the flipping of the ferroelectric order is connected by different symmetries, which makes them have different effects on phonon chirality. The bilayer system can only flip the phonon chirality in the $x$ and $y$ directions, but the four-layer system can flip the phonon chirality in all three directions. For the phonon Berry curvature, its response to interlayer sliding is similar to that of phonon polarization. The influence of interlayer sliding on phonon chirality and Berry curvature brings other related effects; for example, sliding can control the direction of total phonon angular momentum and thermally induced magnetization under a temperature gradient. It can also flip the phonon PAM at the high-symmetry point and the phonon Hall effect under a magnetic field.

\begin{suppinfo}
The first-principles calculations details, stacking ways, results of trilayer PtSe$_2$, symmetric connection, phonon PAM of bilayer and four-layer MoS$_2$ and the calculated phonon Berry curvature.	
\end{suppinfo}

\begin{acknowledgement}
This work was supported by the National Natural Science Foundation of China (Grants Nos.12247149, 11975125, 11890703) and China Postdoctoral Science Foundation 2023M733410. Hao Chen thanks Yan Liang and Yandong Ma for the helpful discussions.
\end{acknowledgement}

\bibliography{achemso-main}

\providecommand{\latin}[1]{#1}
\makeatletter
\providecommand{\doi}
  {\begingroup\let\do\@makeother\dospecials
  \catcode`\{=1 \catcode`\}=2 \doi@aux}
\providecommand{\doi@aux}[1]{\endgroup\texttt{#1}}
\makeatother
\providecommand*\mcitethebibliography{\thebibliography}
\csname @ifundefined\endcsname{endmcitethebibliography}
  {\let\endmcitethebibliography\endthebibliography}{}
\begin{mcitethebibliography}{34}
\providecommand*\natexlab[1]{#1}
\providecommand*\mciteSetBstSublistMode[1]{}
\providecommand*\mciteSetBstMaxWidthForm[2]{}
\providecommand*\mciteBstWouldAddEndPuncttrue
  {\def\EndOfBibitem{\unskip.}}
\providecommand*\mciteBstWouldAddEndPunctfalse
  {\let\EndOfBibitem\relax}
\providecommand*\mciteSetBstMidEndSepPunct[3]{}
\providecommand*\mciteSetBstSublistLabelBeginEnd[3]{}
\providecommand*\EndOfBibitem{}
\mciteSetBstSublistMode{f}
\mciteSetBstMaxWidthForm{subitem}{(\alph{mcitesubitemcount})}
\mciteSetBstSublistLabelBeginEnd
  {\mcitemaxwidthsubitemform\space}
  {\relax}
  {\relax}

\bibitem[Zhang and Niu(2015)Zhang, and Niu]{zhang2015chiral}
Zhang,~L.; Niu,~Q. Chiral phonons at high-symmetry points in monolayer
  hexagonal lattices. \emph{Physical review letters} \textbf{2015}, \emph{115},
  115502\relax
\mciteBstWouldAddEndPuncttrue
\mciteSetBstMidEndSepPunct{\mcitedefaultmidpunct}
{\mcitedefaultendpunct}{\mcitedefaultseppunct}\relax
\EndOfBibitem
\bibitem[Zhu \latin{et~al.}(2018)Zhu, Yi, Li, Xiao, Zhang, Yang, Kaindl, Li,
  Wang, and Zhang]{zhu2018observation}
Zhu,~H.; Yi,~J.; Li,~M.-Y.; Xiao,~J.; Zhang,~L.; Yang,~C.-W.; Kaindl,~R.~A.;
  Li,~L.-J.; Wang,~Y.; Zhang,~X. Observation of chiral phonons. \emph{Science}
  \textbf{2018}, \emph{359}, 579--582\relax
\mciteBstWouldAddEndPuncttrue
\mciteSetBstMidEndSepPunct{\mcitedefaultmidpunct}
{\mcitedefaultendpunct}{\mcitedefaultseppunct}\relax
\EndOfBibitem
\bibitem[Chen \latin{et~al.}(2019)Chen, Lu, Dubey, Yao, Liu, Wang, Xiong,
  Zhang, and Srivastava]{chen2019entanglement}
Chen,~X.; Lu,~X.; Dubey,~S.; Yao,~Q.; Liu,~S.; Wang,~X.; Xiong,~Q.; Zhang,~L.;
  Srivastava,~A. Entanglement of single-photons and chiral phonons in
  atomically thin WSe2. \emph{Nature Physics} \textbf{2019}, \emph{15},
  221--227\relax
\mciteBstWouldAddEndPuncttrue
\mciteSetBstMidEndSepPunct{\mcitedefaultmidpunct}
{\mcitedefaultendpunct}{\mcitedefaultseppunct}\relax
\EndOfBibitem
\bibitem[Li \latin{et~al.}(2019)Li, Wang, Jin, Lu, Lian, Meng, Blei, Gao,
  Taniguchi, Watanabe, \latin{et~al.} others]{li2019momentum}
Li,~Z.; Wang,~T.; Jin,~C.; Lu,~Z.; Lian,~Z.; Meng,~Y.; Blei,~M.; Gao,~M.;
  Taniguchi,~T.; Watanabe,~K., \latin{et~al.}  Momentum-dark intervalley
  exciton in monolayer tungsten diselenide brightened via chiral phonon.
  \emph{ACS nano} \textbf{2019}, \emph{13}, 14107--14113\relax
\mciteBstWouldAddEndPuncttrue
\mciteSetBstMidEndSepPunct{\mcitedefaultmidpunct}
{\mcitedefaultendpunct}{\mcitedefaultseppunct}\relax
\EndOfBibitem
\bibitem[Li \latin{et~al.}(2019)Li, Wang, Jin, Lu, Lian, Meng, Blei, Gao,
  Taniguchi, Watanabe, \latin{et~al.} others]{li2019emerging}
Li,~Z.; Wang,~T.; Jin,~C.; Lu,~Z.; Lian,~Z.; Meng,~Y.; Blei,~M.; Gao,~S.;
  Taniguchi,~T.; Watanabe,~K., \latin{et~al.}  Emerging photoluminescence from
  the dark-exciton phonon replica in monolayer WSe2. \emph{Nature
  communications} \textbf{2019}, \emph{10}, 2469\relax
\mciteBstWouldAddEndPuncttrue
\mciteSetBstMidEndSepPunct{\mcitedefaultmidpunct}
{\mcitedefaultendpunct}{\mcitedefaultseppunct}\relax
\EndOfBibitem
\bibitem[Liu \latin{et~al.}(2020)Liu, van Baren, Liang, Taniguchi, Watanabe,
  Gabor, Chang, and Lui]{liu2020multipath}
Liu,~E.; van Baren,~J.; Liang,~C.-T.; Taniguchi,~T.; Watanabe,~K.;
  Gabor,~N.~M.; Chang,~Y.-C.; Lui,~C.~H. Multipath optical recombination of
  intervalley dark excitons and trions in monolayer WSe 2. \emph{Physical
  Review Letters} \textbf{2020}, \emph{124}, 196802\relax
\mciteBstWouldAddEndPuncttrue
\mciteSetBstMidEndSepPunct{\mcitedefaultmidpunct}
{\mcitedefaultendpunct}{\mcitedefaultseppunct}\relax
\EndOfBibitem
\bibitem[Liu \latin{et~al.}(2021)Liu, Yi, Yang, Lin, Zhang, Zhang, Li, Wang,
  Lee, Tian, \latin{et~al.} others]{liu2021nonlinear}
Liu,~X.; Yi,~J.; Yang,~S.; Lin,~E.-C.; Zhang,~Y.-J.; Zhang,~P.; Li,~J.-F.;
  Wang,~Y.; Lee,~Y.-H.; Tian,~Z.-Q., \latin{et~al.}  Nonlinear valley phonon
  scattering under the strong coupling regime. \emph{Nature materials}
  \textbf{2021}, \emph{20}, 1210--1215\relax
\mciteBstWouldAddEndPuncttrue
\mciteSetBstMidEndSepPunct{\mcitedefaultmidpunct}
{\mcitedefaultendpunct}{\mcitedefaultseppunct}\relax
\EndOfBibitem
\bibitem[Pan and Caruso(2023)Pan, and Caruso]{pan2023vibrational}
Pan,~Y.; Caruso,~F. Vibrational dichroism of chiral valley phonons. \emph{Nano
  Letters} \textbf{2023}, \emph{23}, 7463--7469\relax
\mciteBstWouldAddEndPuncttrue
\mciteSetBstMidEndSepPunct{\mcitedefaultmidpunct}
{\mcitedefaultendpunct}{\mcitedefaultseppunct}\relax
\EndOfBibitem
\bibitem[Xu(2013)]{xu2013ferroelectric}
Xu,~Y. \emph{Ferroelectric materials and their applications}; Elsevier,
  2013\relax
\mciteBstWouldAddEndPuncttrue
\mciteSetBstMidEndSepPunct{\mcitedefaultmidpunct}
{\mcitedefaultendpunct}{\mcitedefaultseppunct}\relax
\EndOfBibitem
\bibitem[Martin and Rappe(2016)Martin, and Rappe]{martin2016thin}
Martin,~L.~W.; Rappe,~A.~M. Thin-film ferroelectric materials and their
  applications. \emph{Nature Reviews Materials} \textbf{2016}, \emph{2},
  1--14\relax
\mciteBstWouldAddEndPuncttrue
\mciteSetBstMidEndSepPunct{\mcitedefaultmidpunct}
{\mcitedefaultendpunct}{\mcitedefaultseppunct}\relax
\EndOfBibitem
\bibitem[Ding \latin{et~al.}(2017)Ding, Zhu, Wang, Gao, Xiao, Gu, Zhang, and
  Zhu]{ding2017prediction}
Ding,~W.; Zhu,~J.; Wang,~Z.; Gao,~Y.; Xiao,~D.; Gu,~Y.; Zhang,~Z.; Zhu,~W.
  Prediction of intrinsic two-dimensional ferroelectrics in In2Se3 and other
  III2-VI3 van der Waals materials. \emph{Nature communications} \textbf{2017},
  \emph{8}, 14956\relax
\mciteBstWouldAddEndPuncttrue
\mciteSetBstMidEndSepPunct{\mcitedefaultmidpunct}
{\mcitedefaultendpunct}{\mcitedefaultseppunct}\relax
\EndOfBibitem
\bibitem[Cui \latin{et~al.}(2018)Cui, Hu, Yan, Addiego, Gao, Wang, Wang, Li,
  Cheng, Li, \latin{et~al.} others]{cui2018intercorrelated}
Cui,~C.; Hu,~W.-J.; Yan,~X.; Addiego,~C.; Gao,~W.; Wang,~Y.; Wang,~Z.; Li,~L.;
  Cheng,~Y.; Li,~P., \latin{et~al.}  Intercorrelated in-plane and out-of-plane
  ferroelectricity in ultrathin two-dimensional layered semiconductor In2Se3.
  \emph{Nano letters} \textbf{2018}, \emph{18}, 1253--1258\relax
\mciteBstWouldAddEndPuncttrue
\mciteSetBstMidEndSepPunct{\mcitedefaultmidpunct}
{\mcitedefaultendpunct}{\mcitedefaultseppunct}\relax
\EndOfBibitem
\bibitem[Li and Wu(2017)Li, and Wu]{li2017binary}
Li,~L.; Wu,~M. Binary compound bilayer and multilayer with vertical
  polarizations: two-dimensional ferroelectrics, multiferroics, and
  nanogenerators. \emph{ACS nano} \textbf{2017}, \emph{11}, 6382--6388\relax
\mciteBstWouldAddEndPuncttrue
\mciteSetBstMidEndSepPunct{\mcitedefaultmidpunct}
{\mcitedefaultendpunct}{\mcitedefaultseppunct}\relax
\EndOfBibitem
\bibitem[Wu and Li(2021)Wu, and Li]{wu2021sliding}
Wu,~M.; Li,~J. Sliding ferroelectricity in 2D van der Waals materials: Related
  physics and future opportunities. \emph{Proceedings of the National Academy
  of Sciences} \textbf{2021}, \emph{118}, e2115703118\relax
\mciteBstWouldAddEndPuncttrue
\mciteSetBstMidEndSepPunct{\mcitedefaultmidpunct}
{\mcitedefaultendpunct}{\mcitedefaultseppunct}\relax
\EndOfBibitem
\bibitem[Miao \latin{et~al.}(2022)Miao, Ding, Wang, Shi, Ye, Li, Yao, Dong, and
  Zhang]{miao2022direct}
Miao,~L.-P.; Ding,~N.; Wang,~N.; Shi,~C.; Ye,~H.-Y.; Li,~L.; Yao,~Y.-F.;
  Dong,~S.; Zhang,~Y. Direct observation of geometric and sliding
  ferroelectricity in an amphidynamic crystal. \emph{Nature Materials}
  \textbf{2022}, \emph{21}, 1158--1164\relax
\mciteBstWouldAddEndPuncttrue
\mciteSetBstMidEndSepPunct{\mcitedefaultmidpunct}
{\mcitedefaultendpunct}{\mcitedefaultseppunct}\relax
\EndOfBibitem
\bibitem[Meng \latin{et~al.}(2022)Meng, Wu, Bian, Pan, Dong, Zhao, Chen, Wu,
  Sun, Fu, \latin{et~al.} others]{meng2022sliding}
Meng,~P.; Wu,~Y.; Bian,~R.; Pan,~E.; Dong,~B.; Zhao,~X.; Chen,~J.; Wu,~L.;
  Sun,~Y.; Fu,~Q., \latin{et~al.}  Sliding induced multiple polarization states
  in two-dimensional ferroelectrics. \emph{Nature Communications}
  \textbf{2022}, \emph{13}, 7696\relax
\mciteBstWouldAddEndPuncttrue
\mciteSetBstMidEndSepPunct{\mcitedefaultmidpunct}
{\mcitedefaultendpunct}{\mcitedefaultseppunct}\relax
\EndOfBibitem
\bibitem[Yang and Wu(2023)Yang, and Wu]{yang2023across}
Yang,~L.; Wu,~M. Across-Layer Sliding Ferroelectricity in 2D Heterolayers.
  \emph{Advanced Functional Materials} \textbf{2023}, 2301105\relax
\mciteBstWouldAddEndPuncttrue
\mciteSetBstMidEndSepPunct{\mcitedefaultmidpunct}
{\mcitedefaultendpunct}{\mcitedefaultseppunct}\relax
\EndOfBibitem
\bibitem[Ji \latin{et~al.}(2023)Ji, Yu, Xu, and Xiang]{ji2023general}
Ji,~J.; Yu,~G.; Xu,~C.; Xiang,~H. General Theory for Bilayer Stacking
  Ferroelectricity. \emph{Physical Review Letters} \textbf{2023}, \emph{130},
  146801\relax
\mciteBstWouldAddEndPuncttrue
\mciteSetBstMidEndSepPunct{\mcitedefaultmidpunct}
{\mcitedefaultendpunct}{\mcitedefaultseppunct}\relax
\EndOfBibitem
\bibitem[Yasuda \latin{et~al.}(2021)Yasuda, Wang, Watanabe, Taniguchi, and
  Jarillo-Herrero]{yasuda2021stacking}
Yasuda,~K.; Wang,~X.; Watanabe,~K.; Taniguchi,~T.; Jarillo-Herrero,~P.
  Stacking-engineered ferroelectricity in bilayer boron nitride. \emph{Science}
  \textbf{2021}, \emph{372}, 1458--1462\relax
\mciteBstWouldAddEndPuncttrue
\mciteSetBstMidEndSepPunct{\mcitedefaultmidpunct}
{\mcitedefaultendpunct}{\mcitedefaultseppunct}\relax
\EndOfBibitem
\bibitem[Vizner~Stern \latin{et~al.}(2021)Vizner~Stern, Waschitz, Cao, Nevo,
  Watanabe, Taniguchi, Sela, Urbakh, Hod, and
  Ben~Shalom]{vizner2021interfacial}
Vizner~Stern,~M.; Waschitz,~Y.; Cao,~W.; Nevo,~I.; Watanabe,~K.; Taniguchi,~T.;
  Sela,~E.; Urbakh,~M.; Hod,~O.; Ben~Shalom,~M. Interfacial ferroelectricity by
  van der Waals sliding. \emph{Science} \textbf{2021}, \emph{372},
  1462--1466\relax
\mciteBstWouldAddEndPuncttrue
\mciteSetBstMidEndSepPunct{\mcitedefaultmidpunct}
{\mcitedefaultendpunct}{\mcitedefaultseppunct}\relax
\EndOfBibitem
\bibitem[Wang \latin{et~al.}(2022)Wang, Yasuda, Zhang, Liu, Watanabe,
  Taniguchi, Hone, Fu, and Jarillo-Herrero]{wang2022interfacial}
Wang,~X.; Yasuda,~K.; Zhang,~Y.; Liu,~S.; Watanabe,~K.; Taniguchi,~T.;
  Hone,~J.; Fu,~L.; Jarillo-Herrero,~P. Interfacial ferroelectricity in
  rhombohedral-stacked bilayer transition metal dichalcogenides. \emph{Nature
  nanotechnology} \textbf{2022}, \emph{17}, 367--371\relax
\mciteBstWouldAddEndPuncttrue
\mciteSetBstMidEndSepPunct{\mcitedefaultmidpunct}
{\mcitedefaultendpunct}{\mcitedefaultseppunct}\relax
\EndOfBibitem
\bibitem[Xiao \latin{et~al.}(2022)Xiao, Gao, Jiang, Gan, Zhang, and
  Li]{xiao2022non}
Xiao,~R.-C.; Gao,~Y.; Jiang,~H.; Gan,~W.; Zhang,~C.; Li,~H. Non-synchronous
  bulk photovoltaic effect in two-dimensional interlayer-sliding
  ferroelectrics. \emph{npj Computational Materials} \textbf{2022}, \emph{8},
  138\relax
\mciteBstWouldAddEndPuncttrue
\mciteSetBstMidEndSepPunct{\mcitedefaultmidpunct}
{\mcitedefaultendpunct}{\mcitedefaultseppunct}\relax
\EndOfBibitem
\bibitem[Liang \latin{et~al.}(2021)Liang, Shen, Huang, Dai, and
  Ma]{liang2021intercorrelated}
Liang,~Y.; Shen,~S.; Huang,~B.; Dai,~Y.; Ma,~Y. Intercorrelated ferroelectrics
  in 2D van der Waals materials. \emph{Materials Horizons} \textbf{2021},
  \emph{8}, 1683--1689\relax
\mciteBstWouldAddEndPuncttrue
\mciteSetBstMidEndSepPunct{\mcitedefaultmidpunct}
{\mcitedefaultendpunct}{\mcitedefaultseppunct}\relax
\EndOfBibitem
\bibitem[Hamada \latin{et~al.}(2018)Hamada, Minamitani, Hirayama, and
  Murakami]{hamada2018phonon}
Hamada,~M.; Minamitani,~E.; Hirayama,~M.; Murakami,~S. Phonon angular momentum
  induced by the temperature gradient. \emph{Physical review letters}
  \textbf{2018}, \emph{121}, 175301\relax
\mciteBstWouldAddEndPuncttrue
\mciteSetBstMidEndSepPunct{\mcitedefaultmidpunct}
{\mcitedefaultendpunct}{\mcitedefaultseppunct}\relax
\EndOfBibitem
\bibitem[Kim \latin{et~al.}(2023)Kim, Vetter, Yan, Yang, Wang, Sun, Yang,
  Comstock, Li, Zhou, \latin{et~al.} others]{kim2023chiral}
Kim,~K.; Vetter,~E.; Yan,~L.; Yang,~C.; Wang,~Z.; Sun,~R.; Yang,~Y.;
  Comstock,~A.~H.; Li,~X.; Zhou,~J., \latin{et~al.}  Chiral-phonon-activated
  spin Seebeck effect. \emph{Nature Materials} \textbf{2023}, \emph{22},
  322--328\relax
\mciteBstWouldAddEndPuncttrue
\mciteSetBstMidEndSepPunct{\mcitedefaultmidpunct}
{\mcitedefaultendpunct}{\mcitedefaultseppunct}\relax
\EndOfBibitem
\bibitem[Strohm \latin{et~al.}(2005)Strohm, Rikken, and
  Wyder]{strohm2005phenomenological}
Strohm,~C.; Rikken,~G.; Wyder,~P. Phenomenological evidence for the phonon Hall
  effect. \emph{Physical review letters} \textbf{2005}, \emph{95}, 155901\relax
\mciteBstWouldAddEndPuncttrue
\mciteSetBstMidEndSepPunct{\mcitedefaultmidpunct}
{\mcitedefaultendpunct}{\mcitedefaultseppunct}\relax
\EndOfBibitem
\bibitem[Sun \latin{et~al.}(2021)Sun, Gao, and Wang]{PhysRevB.103.214301}
Sun,~K.; Gao,~Z.; Wang,~J.-S. Phonon Hall effect with first-principles
  calculations. \emph{Phys. Rev. B} \textbf{2021}, \emph{103}, 214301\relax
\mciteBstWouldAddEndPuncttrue
\mciteSetBstMidEndSepPunct{\mcitedefaultmidpunct}
{\mcitedefaultendpunct}{\mcitedefaultseppunct}\relax
\EndOfBibitem
\bibitem[Werthamer(1970)]{werthamer1970self}
Werthamer,~N. Self-consistent phonon formulation of anharmonic lattice
  dynamics. \emph{Physical Review B} \textbf{1970}, \emph{1}, 572\relax
\mciteBstWouldAddEndPuncttrue
\mciteSetBstMidEndSepPunct{\mcitedefaultmidpunct}
{\mcitedefaultendpunct}{\mcitedefaultseppunct}\relax
\EndOfBibitem
\bibitem[Souvatzis \latin{et~al.}(2008)Souvatzis, Eriksson, Katsnelson, and
  Rudin]{souvatzis2008entropy}
Souvatzis,~P.; Eriksson,~O.; Katsnelson,~M.; Rudin,~S. Entropy driven
  stabilization of energetically unstable crystal structures explained from
  first principles theory. \emph{Physical review letters} \textbf{2008},
  \emph{100}, 095901\relax
\mciteBstWouldAddEndPuncttrue
\mciteSetBstMidEndSepPunct{\mcitedefaultmidpunct}
{\mcitedefaultendpunct}{\mcitedefaultseppunct}\relax
\EndOfBibitem
\bibitem[Errea \latin{et~al.}(2014)Errea, Calandra, and
  Mauri]{errea2014anharmonic}
Errea,~I.; Calandra,~M.; Mauri,~F. Anharmonic free energies and phonon
  dispersions from the stochastic self-consistent harmonic approximation:
  Application to platinum and palladium hydrides. \emph{Physical Review B}
  \textbf{2014}, \emph{89}, 064302\relax
\mciteBstWouldAddEndPuncttrue
\mciteSetBstMidEndSepPunct{\mcitedefaultmidpunct}
{\mcitedefaultendpunct}{\mcitedefaultseppunct}\relax
\EndOfBibitem
\bibitem[Tadano and Tsuneyuki(2015)Tadano, and Tsuneyuki]{tadano2015self}
Tadano,~T.; Tsuneyuki,~S. Self-consistent phonon calculations of lattice
  dynamical properties in cubic SrTiO 3 with first-principles anharmonic force
  constants. \emph{Physical Review B} \textbf{2015}, \emph{92}, 054301\relax
\mciteBstWouldAddEndPuncttrue
\mciteSetBstMidEndSepPunct{\mcitedefaultmidpunct}
{\mcitedefaultendpunct}{\mcitedefaultseppunct}\relax
\EndOfBibitem
\bibitem[Qin \latin{et~al.}(2012)Qin, Zhou, and Shi]{qin2012berry}
Qin,~T.; Zhou,~J.; Shi,~J. Berry curvature and the phonon Hall effect.
  \emph{Physical Review B} \textbf{2012}, \emph{86}, 104305\relax
\mciteBstWouldAddEndPuncttrue
\mciteSetBstMidEndSepPunct{\mcitedefaultmidpunct}
{\mcitedefaultendpunct}{\mcitedefaultseppunct}\relax
\EndOfBibitem
\bibitem[Zhang(2016)]{zhang2016berry}
Zhang,~L. Berry curvature and various thermal Hall effects. \emph{New J. Phys.}
  \textbf{2016}, \emph{18}, 103039\relax
\mciteBstWouldAddEndPuncttrue
\mciteSetBstMidEndSepPunct{\mcitedefaultmidpunct}
{\mcitedefaultendpunct}{\mcitedefaultseppunct}\relax
\EndOfBibitem
\end{mcitethebibliography}

\end{document}